\documentclass[12pt]{elsarticle}

\usepackage{pifont}
\usepackage{times,amsmath,color,amssymb,graphicx,epsfig,cite,psfrag,subfigure}
\usepackage{mathrsfs}
\usepackage{amsfonts}

\usepackage{appendix}
\usepackage{epsfig}
\usepackage{amssymb}
\usepackage{pifont}
\usepackage{times,amsmath,color,amssymb,graphicx,epsfig,cite,psfrag,subfigure}
\usepackage{float}
\usepackage{mathrsfs}
\usepackage{soul}
\usepackage{color}
\usepackage{amsfonts}
\usepackage{multirow}

\usepackage{amssymb}

\journal{Elsevier}

\begin{document}

\begin{frontmatter}
\title{Title\tnoteref{label1}}
\tnotetext[label1]{Part of this work has been presented in IFIP SEC
2012 \citep{kang2012fighting}.}
\author{Xin Kang and Yongdong Wu}
\ead{xkang@i2r.a-star.edu.sg, wydong@i2r.a-star.edu.sg.}
\address{Institute for Infocomm Research, 1 Fusionopolis Way,
$\#$21-01 Connexis, Singapore 138632.}

\title{A Trust-based Pollution Attack Prevention Scheme in Peer-to-Peer Streaming Networks}

\begin{abstract}
Nowadays, peer-to-peer (P2P) streaming systems have become a popular
way to deliver multimedia content over the internet due to their low
bandwidth requirement, high video streaming quality, and
flexibility. However, P2P streaming systems are vulnerable to
various attacks, especially pollution attacks, due to their
distributed and dynamically changing infrastructure. In this paper,
by exploring the features of various pollution attacks, we propose a
trust management system tailored for P2P streaming systems. Both
direct trust and indirect trust are taken into consideration when
designing the trust management system. A new direct trust model is
proposed. A dynamic confidence factor that can dynamically adjust
the weight of direct and indirect trust in computing the trust is
also proposed and studied. A novel double-threshold trust
utilization scheme is given. It is shown that the proposed trust
management system is effective in identifying polluters and
preventing them from further sharing of polluted data chunks.
\end{abstract}

\end{frontmatter}

\section{Introduction}
\subsection{Background and Motivation} The past decade has witnessed the rising of large-scale
multimedia social networks, over which millions of users interact
with each other and exchange media contents in a distributed way.
Among all the multimedia social network applications, peer-to-peer
(P2P) streaming is  popular and successful due to its high
scalability, robustness, and satisfactory performance. Currently,
there are two categories of P2P streaming
systems:\textcolor[rgb]{0.00,0.00,0.00}{ \emph{tree-based}
\citep{mesh3,XSu_201009} and \emph{mesh-based} \citep{mesh1,mesh2}.}
In tree-based P2P streaming systems, the media content is encoded
and divided into small chunks by a root node, and is then
distributed to his children nodes. Then, these children nodes
forward the received chunks to their children nodes. The data chunks
are not forwarded any further at the leaf nodes which reside at the
bottom of the tree. In mesh-based P2P streaming systems, the media
content is encoded and divided into small chunks by peers. Each peer
maintains a buffer map announcing available and desirable chunks.
Peers exchange their buffer maps, and then upload or download data
chunks according to their interests. Unlike the tree-based systems,
mesh-based systems do not need to build and maintain a fixed
streaming topology, and thus overcomes the bandwidth bottleneck
problems existing in tree-based streaming systems. Today's most
popular P2P streaming applications, such as PPTV \citep{PPTV},
PPStream \citep{PPStream}, and SopCast \citep{SopCast}, are all
mesh-based streaming systems.

In these P2P streaming networks, peers are assumed to be well
behaved and non-malicious. To the best of our knowledge, few of them
are designed to be resistant to pollution attacks.
However, due to their distributed and dynamically changing
infrastructure, P2P streaming systems are vulnerable to various
attacks, especially pollution attacks. Malicious peers may
intentionally forge data chunks or alter received data chunks, and
make these polluted data chunks available to other peers. Without
the ability to differentiate between malicious peers and good peers,
peers are highly likely to request and forward polluted data chunks,
consequently degrading the performance of the whole system.
Therefore, effective pollution-resistant mechanisms are badly needed
for P2P streaming systems.

\subsection{Related Work}
\textcolor[rgb]{0.00,0.00,0.00}{A number of scholarly work has been
published in literature on the design of pollution-resistant
mechanisms for P2P streaming systems. In \citep{PDhungel_200708}, by
measuring the PPTV streaming system, the authors showed that without
any pollution-resistant mechanisms, the polluted content could
spread through much of the P2P network. Then, the authors proposed
four possible defenses to pollution attack, namely, blacklisting,
traffic encryption, hash verification, and chunk signing. In
\citep{WConner_2007},  the authors presented a framework to secure
P2P media streaming systems from malicious peers by utilizing a
subset of trusted peers to monitor the bandwidth usage of untrusted
peers and throttle the malicious peers in the system.
In \citep{BHu_ICIP2009}, the authors investigated
the scenario that polluters could upload polluted and clean chunks
alternatively to avoid being detected, and a trust management system
was then proposed to defend this kind of pollution attacks.}

\textcolor[rgb]{0.00,0.00,0.00}{On the other hand, trust management
mechanisms have been extensively studied in literature for a wide
range of applications, such as electronics commerce
\citep{DHMcknight_200209,YAtif_200201,DWManchala_1998}, ad-hoc
networks \citep{Pirzada_2004,ZYan_2003,ZLiu_2004}, P2P networks
\citep{SKamvar_200305,KAberer_2001, ASingh_200309,YWang_200309,
LXiong_conf, LXiong_200407,AJsang_200305,SSong_2005,
YLSun_200602,RZhou_200707,Rzhou_200809,WConner_2009}. However, trust
is in nature a complex psychological concept involving a lot of
complex properties, such as uncertainty, fuzziness, asymmetry, and
time attenuation. The methodology used to model the trust has a
significant influence on the performance of the trust management
system. Trust models should be tailored to meet the specific
requirements of different P2P applications. In this paper, by
exploiting the unique features of pollution attacks, we design a
trust management system to defend against various types of pollution
attacks for P2P multimedia streaming systems. Two closely-related
work are \citep{LXiong_200407} and \citep{RZhou_200707}.  In
\citep{LXiong_200407}, the authors developed a fully distributed
trust management system named as PeerTrust. PeerTrust adopts the
public-key infrastructure for securing trust scores and uses overlay
for trust propagation.  In \citep{RZhou_200707}, the authors
proposed PowerTrust, which is a robust and scalable P2P reputation
system. They leverage the power-law feedback characteristics to
build up a distributed reputation ranking system. PowerTrust can
help peers to identify the most reputable peers quickly and
accurately. However, both PeerTrust and PowerTrust adopt a fixed
weight factor to balance the weight of direct and indirect trust,
and use a single-threshold approach to identify dishonest peers.
Most importantly, the trust models and the trust updates schemes
adopted in PeerTrust and PowerTrust are not tailored to fighting
against pollution attacks.}

\subsection{Main Contributions}
 The main contributions of this paper are listed as follows.
\begin{itemize}
\item \textcolor[rgb]{0.00,0.00,0.00}{A theoretic framework on the modeling of trust management systems to fight against pollution attack in P2P streaming systems is proposed and
investigated.}

\item \textcolor[rgb]{0.00,0.00,0.00}{A dynamic confidence factor is proposed to dynamically adjust the
weight of direct and indirect trust in computing the trust, which is
shown to be pretty effective in reducing the negative effects of the
bad-mouthing attack and the collusion attack. Guidelines on how to
deign such a dynamic confidence factor are given, and two specific
designs of the dynamic confidence factor are proposed and
investigated.}

\item \textcolor[rgb]{0.00,0.00,0.00}{A novel approach to model the direct trust is proposed based on the unique features of pollution
attacks. It is rigorously proved that the proposed trust model is
effective in defending against the on-off pollution attack
introduced in Section \ref{Subsec-OnoffAttack}.}

\item  \textcolor[rgb]{0.00,0.00,0.00}{A novel double threshold trust utilization scheme is proposed, which is shown to better
than the conventional single threshold trust utilization approach.}

\item \textcolor[rgb]{0.00,0.00,0.00}{The performance of the proposed trust management system is
investigated under various types of pollution attacks including
bad-mouth attack, persistent attack, on-off attack, and
collaborative attack. It is shown that the proposed trust management
system is effective in defending against these attacks.}
\end{itemize}

The rest of the paper is organized as follows. Section
\ref{Sec-SysDesignOverview} gives an overview of the design of our
trust management mechanism. Section \ref{TrustFramework} describes
the proposed trust management system in detail. In Section
\ref{PotentialAttacks}, the performance of our trust management
system under various types of pollution attacks is analyzed. In
Section \ref{NumericalResults}, several numerical examples are
presented to validate the proposed studies. Finally, Section
\ref{conclusions} concludes the paper.

\section{System Design Overview} \label{Sec-SysDesignOverview}
In this paper, \textcolor[rgb]{0.00,0.00,0.00}{we consider a
mesh-based P2P streaming network \citep{mesh3}--\citep{mesh2}},
where all the peers can serve as the uploader and the downloader at
the same time. In the proposed system, the media content is encoded
and divided into small chunks by peers. Each peer maintains a buffer
map announcing available and desirable chunks. Peers exchange their
buffer maps, and then upload or download data chunks according to
their interests. To defend against various potential attacks that
are commonly seen in existing P2P streaming networks, we introduce a
trust management system into the P2P streaming network. Under the
proposed trust management system, each peer builds up trust records
of other peers based on their previous direct transactions or
recommendations from other peers. We refer to the trust built on
direct interacting experience as direct trust, and refer to the
trust built on recommendations from third party as indirect trust. A
detail description of direct trust and indirect trust is given in
Section \ref{TrustFramework}.

In our trust management system, we assume that there is no central
database to store the trust values of peers. Instead, the trust
values are computed and stored at each peer itself. We assume there
is a trust manager at each peer. One function of the trust manager
is to do real-time trust evaluation. To evaluate the trustworthiness
of a particular peer, a peer's trust manager sends out the enquiries
on the trust values of the target peer to the peers that have direct
transactions of both the peer and the target peer. Then, the trust
manager computes the trust value of the target peer by doing a
weighted sum of the direct trust and indirect trust values. Another
important function of the trust manager is to feedback submission.
It is responsible for providing recommendations on target peers when
it receives trust value enquires on target peers from other peers.
The benefit of this assumption is 
that the proposed trust management mechanism is fully distributed
and does not rely on a centralized server. Thus, it can be readily
applied to P2P multimedia streaming systems that have distributed structures. 

\begin{figure}[t]
        \centering
        \includegraphics*[width=9cm]{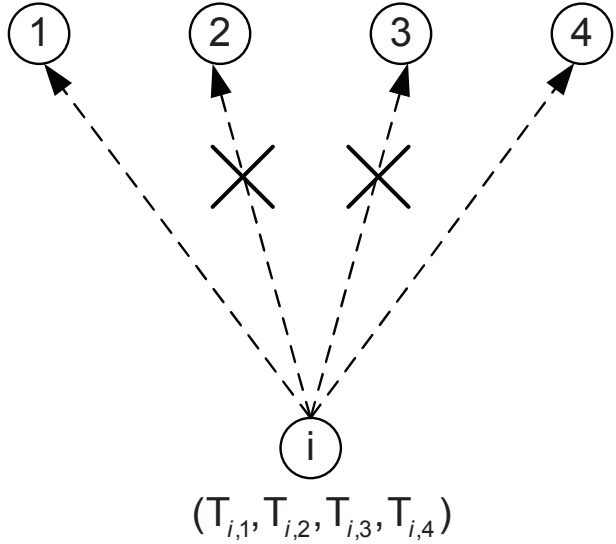}
        \caption{System Model}
        \label{Fig-systemmodel}
\end{figure}

In the proposed trust management system, we use $T_{i,j}(t)$ to
denote the trust that user $i$ has on user $j$ at time $t$. A higher
value of $T_{i,j}$ indicates that user $i$ has a stronger belief
that user $j$ is trustworthy. The trust values are then used by the
peer to decide whether to interact with another peer or not. A peer
will only send its data request to the top $K$ peers from all the
peers having its desired data chunks based on their trust values.
Through this way, peers can reduce the possibility of exposing
themselves to malicious peers, and thus can protect themselves from
potential pollution attacks. This is illustrated in Fig.
\ref{Fig-systemmodel}. Suppose that all the four peers claim that
they have the data chunks that peer $i$ needs, and the trust values
of peer $2$ and $3$ at peer $i$ are low, the trust values of peer
$1$ and $4$ at peer $i$ are high. Then, peer $i$ will only send data
request to peer $1$ and $4$ that are trustworthy.


Detail descriptions of the design of the trust management system is
given in the following section.

\begin{table}[!h]
\tabcolsep 0pt \caption{Notations and Definitions of Basic Terms} 
\begin{center}
\begin{tabular}{|c|c|}
 \hline
\textbf{Notation} & \textbf{Basic Definition}\\
 \hline
 $T_{i,j}$ & Trust value of $j$ evaluated by $i$\\
 \hline
 $D_{i,j}$ & Direct trust value of $j$ evaluated by $i$\\
 \hline
 $I_{i,j}$ & Indirect trust value of $j$ evaluated by $i$\\
 \hline
 $\alpha_{i,j}$ & Dynamic confidence factor of $i$ on $j$\\
 \hline
 $N_{i,j}$ & Number of transactions between $i$ and $j$\\
 \hline
 $N^c_{i,j}$ & Number of clean chunks that $j$ sent to $i$\\
 \hline
 $N^p_{i,j}$ & Number of polluted chunks that $j$ sent to $i$\\
 \hline
 $S_{i,j}$ & Set of peers that have
transactions with $i$,~$j$\\
 \hline
 $C_{i,k}$ & Credibility of $k$ evaluated by $i$\\
 \hline
 $R_{k,j}$ & $k$'s
recommendation value on $j$\\
 \hline
  $\lambda$ & Forgetting factor\\
 \hline
  $\mu$ & Forgiving factor\\
 \hline
  $\theta_i^P$ & Threshold for detecting malicious peers\\
 \hline
  $\theta_i^G$ & Threshold for detecting good peers\\
 \hline
 \end{tabular}
 \end{center}
 \end{table}

\section{Trust Management in P2P Streaming Networks} \label{TrustFramework}

In our trust management system, we use $T_{i,j}(t)$  to denote the
trust that user $i$ has on user $j$ at time $t$. The value of
$T_{i,j}$ is within the range $[0,1]$, with ``$0$'' denoting
distrust and ``$1$'' denoting fully trust.  A higher value of
$T_{i,j}$ indicates that user $i$ has a stronger belief that user
$j$ will upload clean chunks.

Let $D_{i,j}(t)$ and $I_{i,j}(t)$ denote the direct trust and
indirect trust that user $i$ has on user $j$ at time $t$,
$T_{i,j}(t)$ can then be computed as follows
\begin{align}\label{Eq-trust}
T_{i,j}(t)=\alpha_{i,j} D_{i,j}(t)+ (1-\alpha_{i,j} )I_{i,j}(t),
\end{align}
where $0\le\alpha_{i,j}\le1$ is a parameter reflecting user $i$'s
confidence of its direct trust over user $j$. A larger value of
$\alpha_{i,j}$ indicates that user $i$ is more confident of its own
judgement of user $j$, while a smaller value of $\alpha_{i,j}$
indicates that user $i$ relies more on other peers' recommendation
on user $j$.

In the following subsections, we give an in-depth description of
each essential component of the proposed trust management system,
which includes the dynamic confidence factor, the direct trust
model, the indirect trust model, ways to update the trust values,
and how to utilize the trust values. The notations used in this
paper are summarized in Table I.

\subsection{Dynamic Confidence Factor}
Different from the existing literatures (such as
\citep{BHu_ICIP2009}) that use a constant to adjust the weight
between the direct trust and the indirect trust, in this paper, we
define a \emph{dynamic confidence factor} $\alpha_{i,j}$, which is
given as
\begin{align}
\alpha_{i,j}=f\left(N_{i,j}^T\right),
\end{align}
where $f(\cdot)$ is a function, and $N_{i,j}^T$ denotes the number
of direct transactions that has been made between user $i$ and user
$j$ at time $T$. For notation convenience, we drop $t$ in the
discussion.


Basically, $f(\cdot)$ should have the following properties:
\begin{itemize}
\item $\forall N_{i,j}^T \in [0, +\infty)$, $f(N_{i,j}^T)\in [0,1]$.
\item $f(0)=0$, and
$\lim_{N_{i,j}^T\rightarrow\infty}f(N_{i,j}^T)=1$.
\item $f(N_{i,j}^T)$ is a monotonic increasing function of
$N_{i,j}^T$.
\end{itemize}

\emph{Remark:} (a). The first property guarantees that the value of
the trust defined in Equation \eqref{Eq-trust} falls within the
range $[0,1]$. (b). The second property captures the fact that when
there is no direct transaction between user $i$ and user $j$ (i.e.,
$N_{i,j}^T=0$), user $i$ can only rely on the indirect trust values
gathered from other peers to determine its trust of user $j$ (i.e.,
$\alpha_{i,j}=0$). When the number of direct transactions between
user $i$ and user $j$ is sufficiently large, user $i$ can ignore the
indirect trust. (c). The third property captures the fact that the
confidence of user $i$ on its own judgement of the trustworthy of
user $j$ increases when the number of direct transactions between
them increases. (d). It is observed that these properties of
$f(\cdot)$ are similar to those of cumulative distribution functions
(CDF) of random variables \citep{APapoulis_RandomProcess}.
Therefore, the design of $f(\cdot)$ can borrow ideas from the
probability theory.

In this paper, we propose two schemes that satisfy all the
properties mentioned above to design the confidence factor
$\alpha_{i,j}$. The two designs are given as follows.

\underline{\emph{Confidence Factor Design A (CFDA):}}
\begin{align}\label{Eq-Alpha-A}
\alpha_{i,j}=\frac{N_{i,j}^T}{N_{i,j}^T+c},
\end{align}
where $c$ is a positive constant. The value of $c$ has a significant
impact on $\alpha_{i,j}$. For the same $N_{i,j}^T$, a larger $c$
will result in a smaller $\alpha_{i,j}$, while a smaller $c$ will
lead to a larger $\alpha_{i,j}$. In practice, $c$ can be designed as
a tunable parameter that can be tuned by users. This is due to the
fact that different peers have different characteristics. Some peers
are aggressive, and some peers are conservative. For aggressive
peers, they tend to be confidence with their own judgement after a
few transactions, and thus they can set a small value for $c$.  For
conservative peers, they need more transactions to build up the
confidence, and thus they can set a large value for $c$. 


\underline{\emph{Confidence Factor Design B (CFDB):}}
\begin{align}\label{Eq-Alpha-B}
\alpha_{i,j}=1-\beta^{N_{i,j}^T},
\end{align}
where $0<\beta<1$ is a constant. The value of $\beta$ significantly
affects the increasing rate of $\alpha_{i,j}$. For the same
$N_{i,j}^T$, a larger $\beta$ results in a smaller $\alpha_{i,j}$,
while a smaller $\beta$ leads to a larger $\alpha_{i,j}$.  Similar
as CFDA,  $\beta$ should be designed as a tunable parameter that can
be tuned by users. For aggressive peers, they can set a small value
for $\beta$; while for conservative peers, they can set a large
value for $\beta$.


\subsection{Direct Trust}
Direct trust is the trust of a peer on another peer based on their
direct interacting experience. It is established only based on
previous direct transactions between peers. In a P2P streaming
system, it is usually determined by two variables: the number of
received clean chunks and the number of received polluted chunks.
Let $N^c_{i,j}(t)$ and $N^p_{i,j}(t)$ denote the total number of
clean chunks and polluted chunks that user $i$ has received from
user $j$ at time $t$, the direct trust $D_{i,j}(t)$ that user $i$
has on user $j$ at time $t$ can be defined as
\begin{align}
D_{i,j}(t)=g\left(N^c_{i,j}(t),N^p_{i,j}(t)\right),
\end{align}
where $g(\cdot,\cdot)$ is a two-dimensional function. Basically,
$g(\cdot,\cdot)$ should have the following properties:
\begin{itemize}
\item $\forall N_{i,j}^c,  N_{i,j}^p\in [0, +\infty)$, $g\left(N^c_{i,j},N^p_{i,j}\right)\in [0,1]$.
\item $g\left(N^c_{i,j},N^p_{i,j}\right)$ is an increasing function of
$N_{i,j}^c$, and is a decreasing function of $N_{i,j}^p$.
\end{itemize}

In fact, there are already several direct trust models exiting in
the literature. In the following, we list two prevalent direct trust
models.

\underline{\emph{Direct Trust Model A (DTMA):}}
\begin{align}\label{Eq-Directrust-A}
D_{i,j}(t)=\frac{N^c_{i,j}(t)}{N^c_{i,j}(t)+N^p_{i,j}(t)}.
\end{align}
\textcolor[rgb]{0.00,0.00,0.00}{This model has been used in
\citep{BHu_ICIP2009} and \citep{YLSun_200602}}. It represents the
ratio of the number of clean chunks vs the total number of chunks
that user $i$ has received from user $j$. Another model is

\underline{\emph{Direct Trust Model B (DTMB):}}
\begin{align}\label{Eq-Directrust-C}
D_{i,j}(t)=\frac{N^c_{i,j}(t)+1}{N^c_{i,j}(t)+N^p_{i,j}(t)+2}.
\end{align}
\textcolor[rgb]{0.00,0.00,0.00}{This model has been used in
\citep{AJsang_200305} and \citep{YLSun_200602}.} It is established
based on beta-function.

It is observed that if a malicious peer sends clean and polluted
chunks alternatively to the peers that request data from it, it can
easily keep its trust value above certain threshold if DTMA or DTMB
is adopted. For example, if the malicious peer performs the
pollution attack by sending one polluted data chunk after sending
every two clean data chunks, it can keep its trust value above
$0.5$. In this way, it can avoid not being detected as a polluter,
and keep sending polluted data chunks to the victims. This type of
attack is referred to as \emph{on-off attack}. This indicates that
DTMA and DTMB are vulnerable to the on-off attack. They cannot be
used alone, and must be used together with other techniques to fight
against the on-off attack. This inevitably increases the complexity
and difficulty of the system design. In this paper, we propose a
novel way to model the direct trust, which is resistant to the
on-off attack.

\underline{\emph{Proposed Direct Trust Model (PDTM):}}
\begin{align}\label{Eq-Directrust-B}
D_{i,j}(t)=e^{-\rho
N^p_{i,j}(t)}\frac{N^c_{i,j}(t)}{N^c_{i,j}(t)+\eta},
\end{align}
where $\rho$ and $\eta$ are positive constants, and $e^{(\cdot)}$ is
the exponential function. It is easy to verify that the value of
$D_{i,j}$ is within the range $[0,1]$, and $D_{i,j}$ is an
increasing function with regard to $N_{i,j}^c$ and a decreasing
function with regard to $N_{i,j}^p$. The value of the two parameters
$\rho$ and $\eta$ has a great impact on the value of $D_{i,j}(t)$.
Therefore, how to set the value of $\rho$ and $\eta$ is of great
importance to the performance of PDTM. PDTM is rigorously proved to
be resistant to the on-off attack when $\rho$ and $\eta$ satisfy the
condition $\rho>\ln(1+\frac{1}{\eta})$. The details and the proof
are given in Section \ref{Subsec-OnoffAttack}.


\subsection{Indirect Trust}
Indirect trust is the trust of a peer on another peer obtained via
third-party peers' recommendations. Indirect trust is very important
when two peers have little or no direct interactions. Indirect trust
is established through trust propagation, i.e., trustworthy peers
are more likely to give honest feedbacks than distrusted peers.
Usually, indirect trust is determined by two key factors: the
credibility of the third-party peer and its recommendation value of
the trustee.

\textcolor[rgb]{0.00,0.00,0.00}{Similar to  \citep{BHu_ICIP2009}, in
this paper, we define the indirect trust as}
\begin{align}\label{Eq-Indirectrust}
I_{i,j}(t)\triangleq\frac{\sum_{k \in S_{i,j}(t)} C_{i,k}(t)
R_{k,j}(t)}{\sum_{k \in S_{i,j}(t)} C_{i,k}(t)},
\end{align}
where $S_{i,j}(t)$ denotes the set of peers that have direct
transactions with both peer $i$ and peer $j$. $C_{i,k}(t)$ is the
credibility of peer $k$, and $R_{k,j}(t)$ is user $k$'s
recommendation value of user $j$ based on their interaction
experience.

In this paper, we design $C_{i,k}(t)$ and $R_{k,j}(t)$ as follows
\begin{align}\label{Eq-Cre}
C_{i,k}(t)&=D_{i,k}(t),\\
R_{k,j}(t)&=D_{k,j}(t),\label{Eq-Recon}
\end{align} where $D_{i,k}(t)$ is
the peer $i$'s direct trust on peer $k$, and $D_{k,j}(t)$ is the
peer $k$'s direct trust on peer $j$. It is observed that $k$'s
recommendation on peer $j$ is weighted proportionally by its own
credibility. This design has two advantages. First, the value of
peer $k$' recommendation on peer $j$ can not be larger than its
credibility. This perfectly emulates human's psychology, i.e., when
a person establishes a trust relationship with another person
(referee) through a recommender, the trust between the person and
the referee is usually not as strong as that between the person and
the recommender. Second, peer $k$' recommendation on peer $j$ must
be based on its direct trust value of peer $j$. In this way, this
makes the indirect trust resilient to malicious peers who can
manipulate their recommendations to cause maximal damage to the
network.

\subsection{Trust Updates}
Intuitively, the recent interactions should have more weight than
old interactions in computing the trust value. In this paper, we
assume that the interactions made within the recent $\Delta t$ time
have the same weight, and the weight of the interactions made older
than $\Delta t$ will experience certain attenuation. Mathematically,
the update functions can be written as
\begin{align}
N^c_{i,j}(t^\prime)=e^{-\lambda\Delta t}
N^c_{i,j}(t)+\left(N^c_{i,j}(t^\prime)-N^c_{i,j}(t)\right),\\
N^p_{i,j}(t^\prime)=e^{-\mu\Delta t}
N^p_{i,j}(t)+\left(N^p_{i,j}(t^\prime)-N^p_{i,j}(t)\right),
\end{align}
where $\lambda$ and $\mu$ are positive constants, and
$t^\prime=t+\Delta t$. In this paper, we refer to $\lambda$ and
$\mu$ as \emph{forgetting factor} and \emph{forgiving factor},
respectively.  We request $\lambda>\mu$, which makes our trust
management system remembers the unpleasant interactions longer than
the pleasant interactions.

We introduce the trust update functions due to the following three
reasons. Firstly, the \emph{forgetting} property provides an
incentive for peers to keep uploading clean chunks to maintain or
increase their trust values. If the trust computation is only
carried out in a cumulative manner without the forgetting property,
a peer will have diminishing incentives to behave honestly when it
has established a high trust value. This is due to the fact that the
negative behaviors will play a little role in changing the peer's
trust value at this time. However, if older trust values is
discounted with time, a peer's recent behavior always matters and
the peer has continuing incentives to behave honestly to maintain or
increase its trust values. Secondly, the \emph{forgiving} property
allows good peers to wipe off their bad transaction records caused
by the bad network conditions. In practice, package loss is
inevitable when the network is congested. This will result in
incomplete data chunks. If a peer receives such kinds of data
chunks, it will treat these data chunks as polluted data chunks and
reduce the trust value of the sender though the sender is innocent.
With the forgiving property, peers will forget these unpleasant
transactions. Finally, the \emph{forgiving} property also gives a
chance for the distrusted peers to rejoin the network after a
sufficient long waiting time during which they may become good. 


\subsection{Utilizations of Trust Values}
With the trust management system introduced in this section, peers
can easily compute the trust values of other peers. The trust values
can then be used by peers to identify polluters, and to determine
whether to perform a transaction with another peer. A conventional
approach is to set up a trust threshold to differentiate polluters
from good peers. For example, peer $i$ can set up a threshold
$\theta_i$. If a peer's trust value is below $\theta_i$, peer $i$
identifies it as a polluter and will not perform any further
transactions with it. On the other hand, if a peer's trust value is
above $\theta_i$, peer $i$ identifies it as a good peer and will
perform the next transaction with it. The value of the threshold can
be different for different peers. This is due to the fact that
different peers may have different perception over the same trust
value. This approach is easy to implement. However, it has some
deficiencies. For example, if peer $i$ sets a high value for
$\theta_i$, it will lose the opportunities to perform transactions
with peers whose low trust values are caused by previous bad network
conditions. On the other hand, if peer $i$ sets a low value for
$\theta_i$, it may make the trust management system vulnerable to
potential pollution attacks. Therefore, in this paper, we propose a
double threshold approach to utilize the trust values.

Suppose peer $i$ decides to make a transaction with peer $j$ with
probability $p_{i,j}(t)$ at time $t$, then $p_{i,j}(t)$ can be
determined by
\begin{align}
p_{i,j}(t)=\left\{\begin{array}{ll}
                 0, & \mbox{if}~T_{i,j}(t)<\theta_i^P, \\
                 \chi_{i,j}, & \mbox{if}~\theta_i^P\le T_{i,j}(t)<\theta_i^G, \\
                 1, & \mbox{if}~T_{i,j}(t)\ge \theta_i^G,
               \end{array}
\right.
\end{align}
where $\theta_i^P$ and $\theta_i^G$ are the thresholds for peer $i$
to identify malicious and good peers, respectively. If the trust
value of peer $j$ is below $\theta_i^P$, peer $i$ will identify it
as a polluter and will not perform any further transactions with it;
if the trust value of peer $j$ is larger than $\theta_i^G$,  peer
$i$ will identify it as a good peer and will perform the next
transaction with it without hesitation. However, if the trust value
of peer $j$ is between $\theta_i^P$ and $\theta_i^G$, it is hard for
peer $i$ to judge whether peer $j$ is a polluter or a good peer
experiencing bad network conditions. In this scenario, peer $i$ will
perform the next transaction with peer $j$ with probability
$\chi_{i,j}$. It is worth pointing out that peer $i$ can set
different $\chi_{i,j}$ for different peer $j$, depending on the
content of the potential transaction. For example, peer $i$ is
willing to set a high value of $\chi_{i,j}$ for a peer $j$ that has
data chunks which are closer to its playback time.


\section{System Performance Analysis under Potential Attacks} \label{PotentialAttacks} In
this subsection, we give an introduction of the commonly seen
attacks in P2P streaming networks,
\textcolor[rgb]{0.00,0.00,0.00}{such as \emph{bad-mouthing attack}
\citep{YLSun_200802}, \emph{persistent attack}
\citep{PDhungel_200708,WConner_2007}, \emph{on-off attack}
\citep{BHu_ICIP2009,YLSun_200602}, and \emph{collaborative attack}
\citep{BHu_ICIP2009,YLSun_200602}.} The performance of the proposed
trust management system are then investigated under these attacks.

\subsection{Bad-Mouthing Attack} \textcolor[rgb]{0.00,0.00,0.00}{\emph{Bad-Mouthing Attack} \citep{YLSun_200802} refers
to the scenario that a single malicious peer or a group of malicious
peers deliberately provides negative recommendations to frame up
good peers.} If there is only one malicious user, the negative
effect of the bad-mouthing attack is quite limited and thus can be
ignored. This is due to the fact that the indirect trust is obtained
from the recommendations of a group of peers, and a single peer's
malicious recommendation is not able to make a big change of the
indirect trust value. However, when a group of malicious peers
collude and give negative recommendations, the value of indirect
trust will be affected.

In our trust management system, the following two ways are adopted
to fight against bad-mouthing attacks.

a). \emph{Filtering out potential malicious recommendations}. When
computing the indirect trust $I_{i,j}(t)$, peer $i$ only select the
top $K$ peers based on the value of $D_{i,j}(t)$ from the set
$S_{i,j}(t)$. By doing this, a peer can effectively avoid the
malicious recommendations from untrustworthy peers. The value of $K$
can be determined by each peer itself based on its own needs.

b). \emph{Reducing the weight of indirect trust}. Bad-mouthing
attacks are unavoidable as long as recommendations are taken into
consideration. Therefore, reducing the weight of indirect trust in
the trust computation is a good way to defend against bad-mouthing
attacks. The proposed two schemes to dynamically adjust the
confidence factor given in equations \eqref{Eq-Alpha-A} and
\eqref{Eq-Alpha-B} can effectively reduce the weight of indirect
trust, and thus increase the trust management system's resistant to
the bad-mouthing attack.

\subsection{Persistent Attack} \textcolor[rgb]{0.00,0.00,0.00}{\emph{Persistent attack} \citep{PDhungel_200708,WConner_2007} refers to
the scenario that a malicious peer keeps sending polluted chunks to
the peers that request data from it.} This kind of attacks is very
easy to handle when the number of malicious peers are not large.
When a malicious peer performs persistent attack, its trust value
decreases fast. When its trust value falls below the predetermined
threshold, it can be easily detected as a polluter, and will be
prevented from further sharing of polluted data. However, if there
are a lot of malicious peers existing in the network, the
conventional trust management system may be not sufficient.  This is
because the trust value of a malicious peer is inversely
proportional to the polluted data chunks it sends out in
conventional trust management systems. Thus, it takes time for the
trust value of a malicious peer to drop below certain threshold. If
a lot of malicious peers attack the victim at the same time, the
victim may be  not  able to survive until it can identify malicious
peers.

The proposed trust management mechanism is effective in handling
with this type of attacks due to the following reasons. First, in
our trust management mechanism, a peer will always send data request
to those peers that have transactions with it before, and select the
top $K$ peers based on the value of $D_{i,j}$ from these peers. In
this way, peers can reduce their exposure to malicious peers.
Secondly, in our trust management system, the trust value drops
exponentially with respect to the number of the polluted data
chunks. As a result, the trust value of malicious peers drops below
the prescribed threshold within a few data chunks, and thus the
victim can identify the these malicious peers quickly.

\subsection{On-Off Attack}\label{Subsec-OnoffAttack} \textcolor[rgb]{0.00,0.00,0.00}{\emph{On-off
attack} \citep{BHu_ICIP2009,YLSun_200602} refers to the scenario
that a malicious peer sends clean and polluted chunks alternatively
to the peers that request data from it. }By doing this, the malicious
peer can keep its trust value above the predetermined threshold, and
thus avoid being identified as a polluter. The on-off attack
exploits the fact that most of the trust management mechanisms are
designed to tolerate certain levels of unintentionally polluted
chunks (such as incomplete data chunks and erroneous data chunks)
due to bad network conditions.

To combat the on-off attack, an effective way is to design a trust
management system in which the dropping rate of trust value is
larger than its increasing rate, i.e., the trust value drops sharply
when the peer uploads polluted chunks, and accumulates slowly when
the peer uploads the same number of clean chunks. If a trust
management mechanism satisfies this condition, we say it is
resistant to on-off attack.

\underline{\emph{Proposition 4.1:}} The proposed direct trust model
given in \eqref{Eq-Directrust-B} 
is resistant to
on-off attack when $\rho>\ln(1+\frac{1}{\eta})$.

\textbf{Proof:} Let the current trust value be $D_{i,j}(t)$. Suppose
peer $j$ continuously uploads $N$ polluted chunks to peer $i$ in the
following transaction, then the trust value drop denoted by $\Delta
D^{de}_{i,j}(t)$ is
\begin{align}\label{Eq-TDrop}
\Delta
&D^{de}_{i,j}(t)=D_{i,j}(t)-e^{-\rho\left(N^p_{i,j}(t)+N\right)}\frac{N^c_{i,j}(t)}{N^c_{i,j}(t)+\eta}\nonumber\\
&=\left(e^{-\rho
N^p_{i,j}(t)}-e^{-\rho\left(N^p_{i,j}(t)+N\right)}\right)\frac{N^c_{i,j}(t)}{N^c_{i,j}(t)+\eta}.
\end{align}
On the other hand, if peer $j$ continuously uploads $N$ clean chunks
to peer $i$ in the following transaction, then the trust value
increase denoted by $\Delta D^{in}_{i,j}(t)$ is
\begin{align}\label{Eq-TIncrease}
\Delta
&D^{in}_{i,j}(t)=e^{-\rho N^p_{i,j}(t)}\frac{N^c_{i,j}(t)+N}{N^c_{i,j}(t)+N+\eta}-D_{i,j}(t)\nonumber\\
&=e^{-\rho N^p_{i,j}(t)}\frac{\eta
N}{\left(N^c_{i,j}(t)+\eta\right)\left(N^c_{i,j}(t)+N+\eta\right)}.
\end{align}
The proposed trust management mechanism is resistant to on-off
attack when $\Delta D^{de}_{i,j}(t)/\Delta D^{in}_{i,j}(t)>1$. With
equations given in \eqref{Eq-TDrop} and \eqref{Eq-TIncrease}, we
have
\begin{align}
\frac{\Delta D^{de}_{i,j}(t)}{\Delta
D^{in}_{i,j}(t)}&=\left(1-e^{-\rho N}\right)\frac{N^c_{i,j}(t)}{\eta
N}\left(N^c_{i,j}(t)+N+\eta\right)\nonumber\\
&\ge\left(1-e^{-\rho N}\right)\left(\eta+N\right).
\end{align}
It is easy to verify that $h(N)\triangleq\left(1-e^{-\rho
N}\right)\left(\eta+N\right)$ is an increasing function with regard
to $N$. Therefore, $h(N)>1, ~\forall N$, if $h(1)>1$, which is
equivalent to $\rho>\ln(1+\frac{1}{\eta})$. $\blacksquare$

\subsection{Collaborative Attack} \textcolor[rgb]{0.00,0.00,0.00}{\emph{Collaborative Attack} \citep{BHu_ICIP2009,YLSun_200602}
refers to the scenario that a group of malicious peers work together
to strategically send polluted data chunks to the target peers.} A
typical scenario is that one or some malicious peers in the group
keep sending polluted data chunks to the target peers, while others
send valid data to gain the trust of the target peers and give high
recommendation trust values on these malicious peers. The proposed
trust management mechanism is quite effective in defending against
this kind of attacks due to the adoption of dynamic confidence
factor. This is due to the fact that the dynamic confidence factor
can effectively increase the weight of direct trust and reduce the
weight of indirect trust with the increasing of the number of
transactions. When the number of transactions exceeds certain level,
the trust value is dominated by direct trust, and the indirect trust
can be ignored.

A more advanced type of collaborative attacks is the scenario that a
group of malicious peers take turns to send polluted data chunks to
the target peers, and at the same time, they give high
recommendations on each other. This type of attack is a little
complicated and is in general not easy to handle under the
conventional trust management mechanisms. However, due to the
adoption of the dynamic confidence factor and the proposed direct
trust model, the proposed trust management mechanism is effective in
fighting against this type of attack. The individual behavior of
each malicious peer in the group is actually same as that of the
on-off attack. The difference part is that these malicious peers
give high recommendation on each other. Through this way, they can
increase the indirect trust values of the attackers, consequently
misleading the victims judgement on the attackers. However, as
mentioned before, the proposed dynamic confidence factor can
effectively reduce the weight of the indirect trust with increasing
of the number of transactions. When the number of transactions
exceeds certain level, the trust value will be dominated by the
direct trust, and the indirect trust can be ignored. On the other
hand, as shown in Section \ref{Subsec-OnoffAttack}, the proposed
direct trust model can effectively prevent the on-off attack.
Therefore, with the combination of these two components, the
proposed trust management mechanism can easily handle this type of
collaborative attacks.

\section{Simulation Results}\label{NumericalResults}
In this section, several examples are provided to evaluate the
performance of the proposed studies. It is shown that the proposed
trust management is quite effective in fighting against various
types of pollution attacks. In the simulations, we consider a
network with 10000 nodes. The network topology is generated by BRITE
\citep{Brite}, and is then imported to NS-2 \citep{NS2} to do
simulation. Without specific declaration, we assume that CFDA with
$c=1$ is adopted to compute the dynamic confidence factor.  We
assume $\eta=1$, and $\rho=\ln(1+\frac{1}{\eta})$ for computing the
direct trust. The detailed simulation setup for each experiment is
clearly described in each individual example studied below.

\subsection{Experiment 1: Constant Confidence Factor vs. Dynamic Confidence Factor}
\begin{figure}[t]
        \centering
        \includegraphics*[width=10cm]{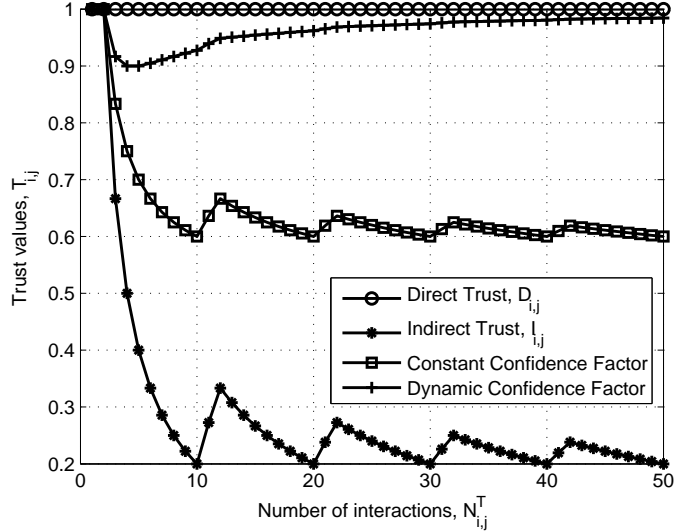}
        \caption{Constant confidence factor vs. dynamic confidence factor}
        \label{Fig-CnvsDyCF}
\end{figure}
In this experiment, we let peer $j$ keep uploading clean chunks to
peer $i$. We assume that peer $j$ is under bad-mouthing attack, and
some malicious peers give bad recommendations on peer $j$ for $80$
percent of total transactions. Peer $i$ computes the trust values of
peer $j$ for $50$ interactions based on the constant confidence
factor scheme (CCFS) and our dynamic confidence factor scheme
(DCFS), respectively. For the CCFS, we assume that the confidence
factor is 0.5. For the DCFS, we adopt the CFDA proposed in Section
\ref{TrustFramework}. For fair comparison, we assume the direct
trust is computed by \eqref{Eq-Directrust-A}. It is seen from Fig.
\ref{Fig-CnvsDyCF}, the values of direct trust is always equal to 1
since peer $j$ keeps uploading clean chunks to peer $i$. While the
values of indirect trust is very low since $j$ is suffering from
bad-mouthing attack. It is observed from Fig. \ref{Fig-CnvsDyCF}
that the trust values of peer $j$ obtained based on CCFS deviate far
from the true trust values, while the values obtained based on DCFS
are quite close to the true trust values. Besides, the difference
between the trust values computed based on DCFS and the the true
trust values diminishes with the increasing of the number of
interactions. This example demonstrates the fact that the proposed
DCFS is quite effective in fighting against bad-mouthing attacks.


\subsection{Experiment 2: Existing Direct Trust Models vs. Proposed Direct Trust Model}
\begin{figure}[!t]
\centering \subfigure[Proposed direct trust model]{\label{Fig-PDTM}
\includegraphics*[width=6.5cm]{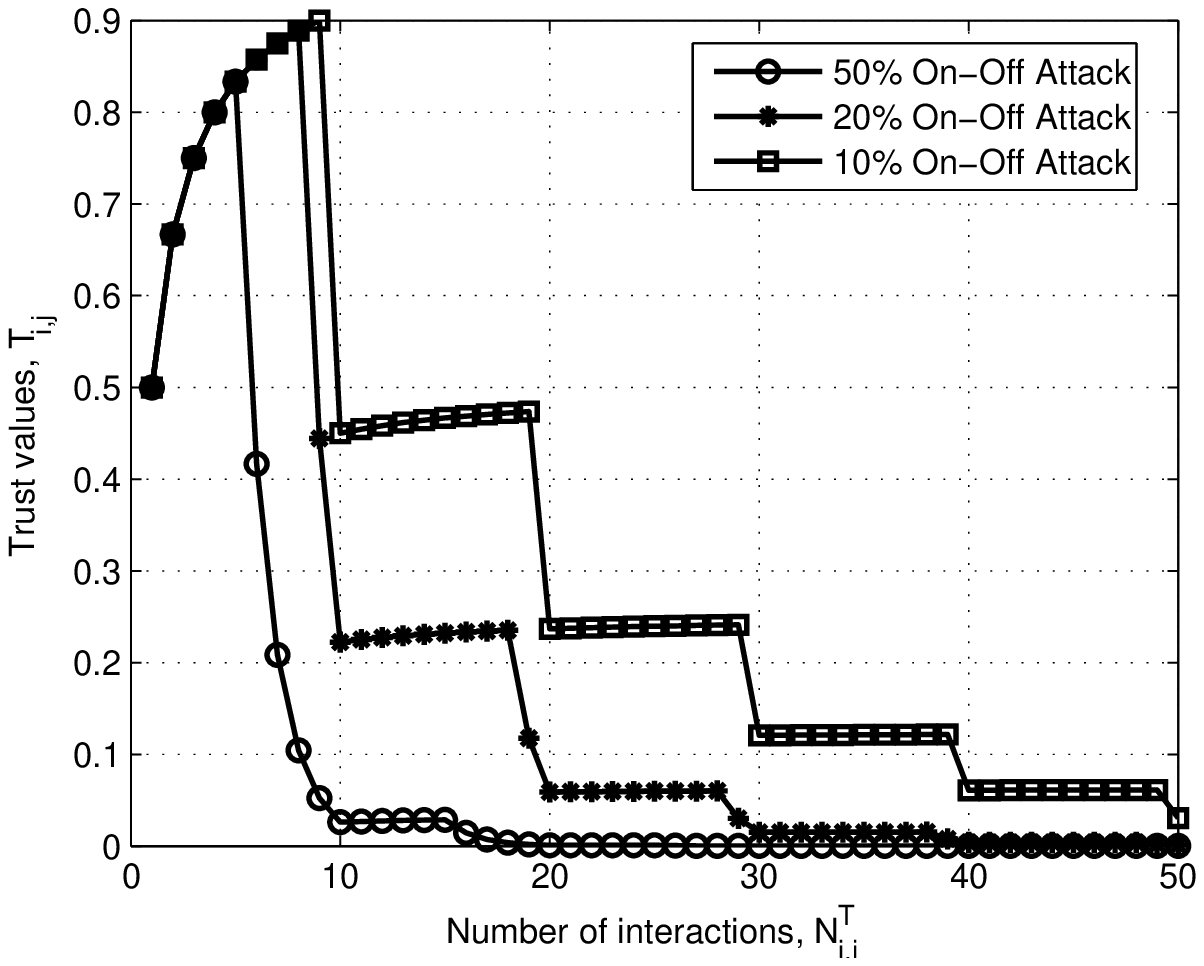}} \hfil
\subfigure[Existing direct trust
model]{\label{Fig-DTMA}\includegraphics*[width=6.5cm]{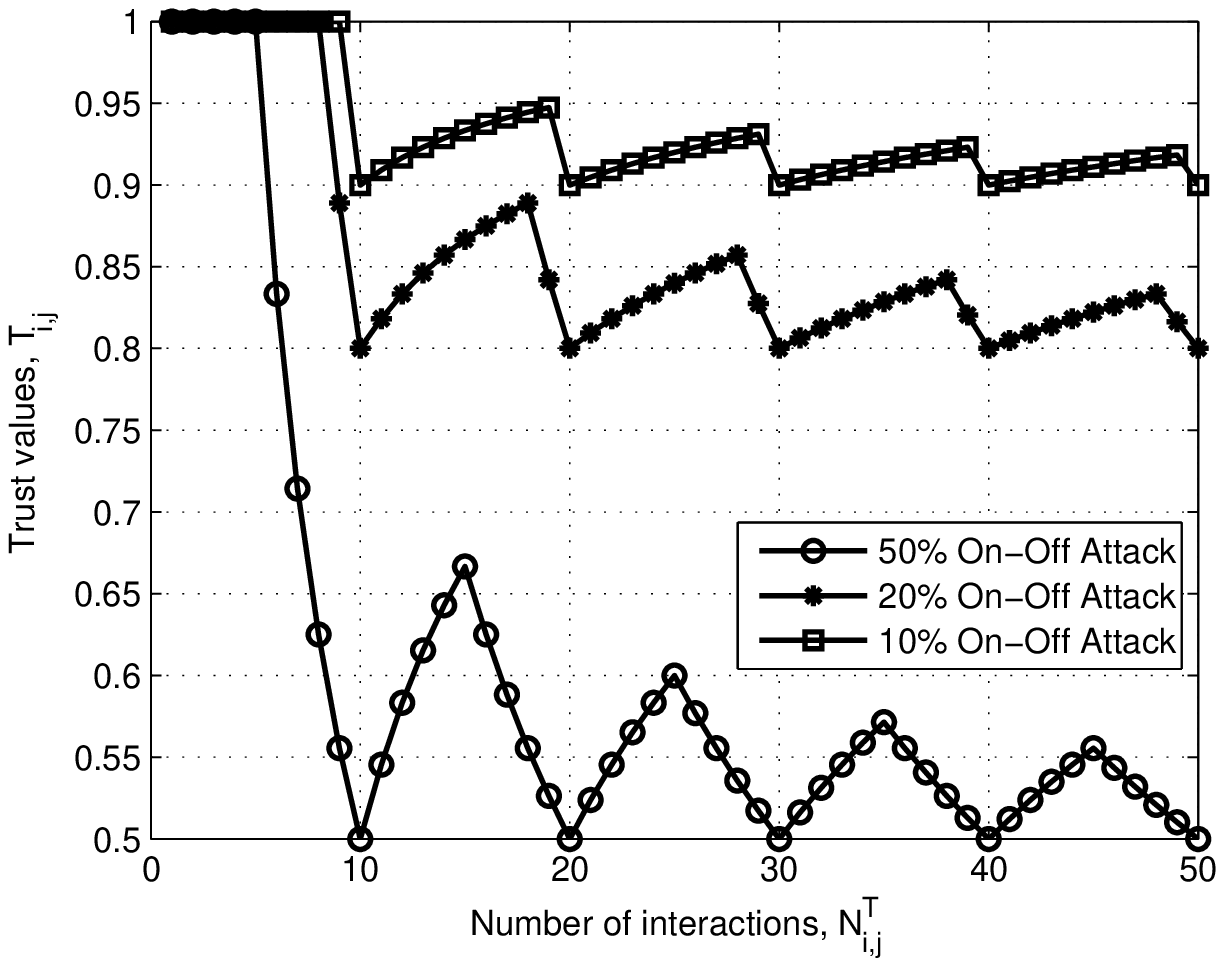}}
\caption{Existing direct trust models vs. proposed direct trust
model}
\end{figure}
In this experiment, we let peer $j$ perform on-off attacks on peer
$i$. Three different on-off ratios ($50\%$, $20\%$, and $10\%$) are
considered. When the attacker is in ``on'' mode, it sends polluted
data chunks to the target peer. When the attacker is in ``off''
mode, it pretends to be a good peer, and sends clean data chunks to
the target peer. The on-off ratio denotes the ratio of the duration
of the ``on'' mode to the duration of the entire cycle. As explained
before, DTMA and DTMB have similar performance. Thus, in this part,
we only compare the performance of the proposed direct trust model
with that of DTMA. The trust values of peer $j$ are computed for
$50$ interactions based on DTMA and PDTM, respectively. It is
observed from Fig. \ref{Fig-PDTM} that the dropping rates are much
larger than the increasing rate under PDTM. Therefore, the trust
values obtained under PDTM are gradually decreasing in the long run.
It is observed that under the $50\%$ on-off attack, the trust value
drops below $0.1$ within $10$ interactions. On the other hand, it is
observed from Fig. \ref{Fig-DTMA} that the trust values computed
under DTMA are maintained above certain thresholds. For example,
under the $20\%$ on-off attack, the trust values of the attacker are
maintained above $0.8$. Even under the $50\%$ on-off attack, the
trust values of the attacker are maintained above $0.5$, which
indicates that DTMA is not resistant to the on-off attack.

\subsection{Experiment 3: Single Trust Threshold Scheme vs. Proposed Double Trust Thresholds Scheme}
\begin{figure}[t]
        \centering
        \includegraphics*[width=10cm]{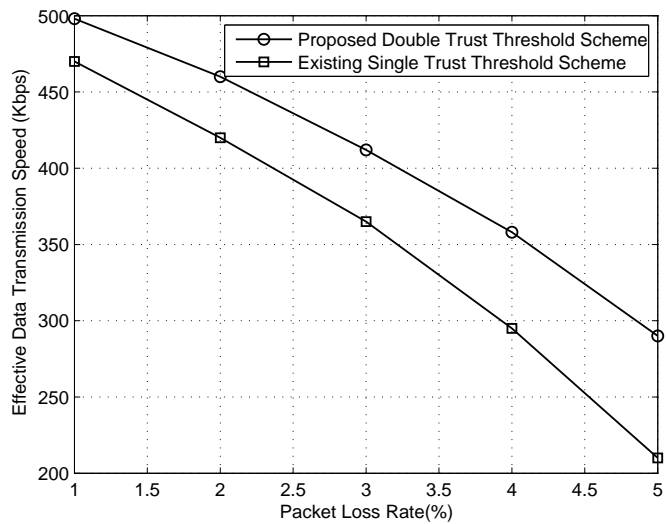}
        \caption{Single Trust Threshold vs. Proposed Double Trust Thresholds}
        \label{Fig-DoubleVsSingle}
\end{figure}
In this experiment, we let $20\%$ of the peers to be malicious.
$10\%$ of them perform persistent attack, and the other $10\%$ of
them perform on-off attack with a $20\%$ on-off ratio. Then, we
observe the performance of peer $i$ under different network
conditions using the two schemes, respectively. For the single trust
threshold scheme, we set the trust threshold of peer $i$ as $0.8$.
For the proposed double trust thresholds scheme, we adopt the
following parameters: $\theta_i^P=0.5$, $\theta_i^G=0.9$, and
$\chi_{i,j}=0.5, \forall j$. It is observed from Fig.
\ref{Fig-DoubleVsSingle} that peers under the proposed scheme can
always achieve higher data rate as compared to the existing single
trust threshold scheme. This is due to the fact that the proposed
scheme reduces the probability of mistaking a good peer experiencing
bad network conditions as a polluter. It is also observed that the
gap between the proposed scheme and the single threshold scheme
increases when the network conditions become worse. This is because
when the network conditions become worse, the trust value of more
peers are affected, and thus these peers are regarded as malicious
peers by the conventional single trust threshold scheme. On the
other side, the proposed scheme allows peer $i$ to perform
transactions with these peers with certain probability, and thus
results in a higher data rate.

\subsection{Experiment 4: Performance under collaborative attack}
\begin{figure}[t]
        \centering
        \includegraphics*[width=10cm]{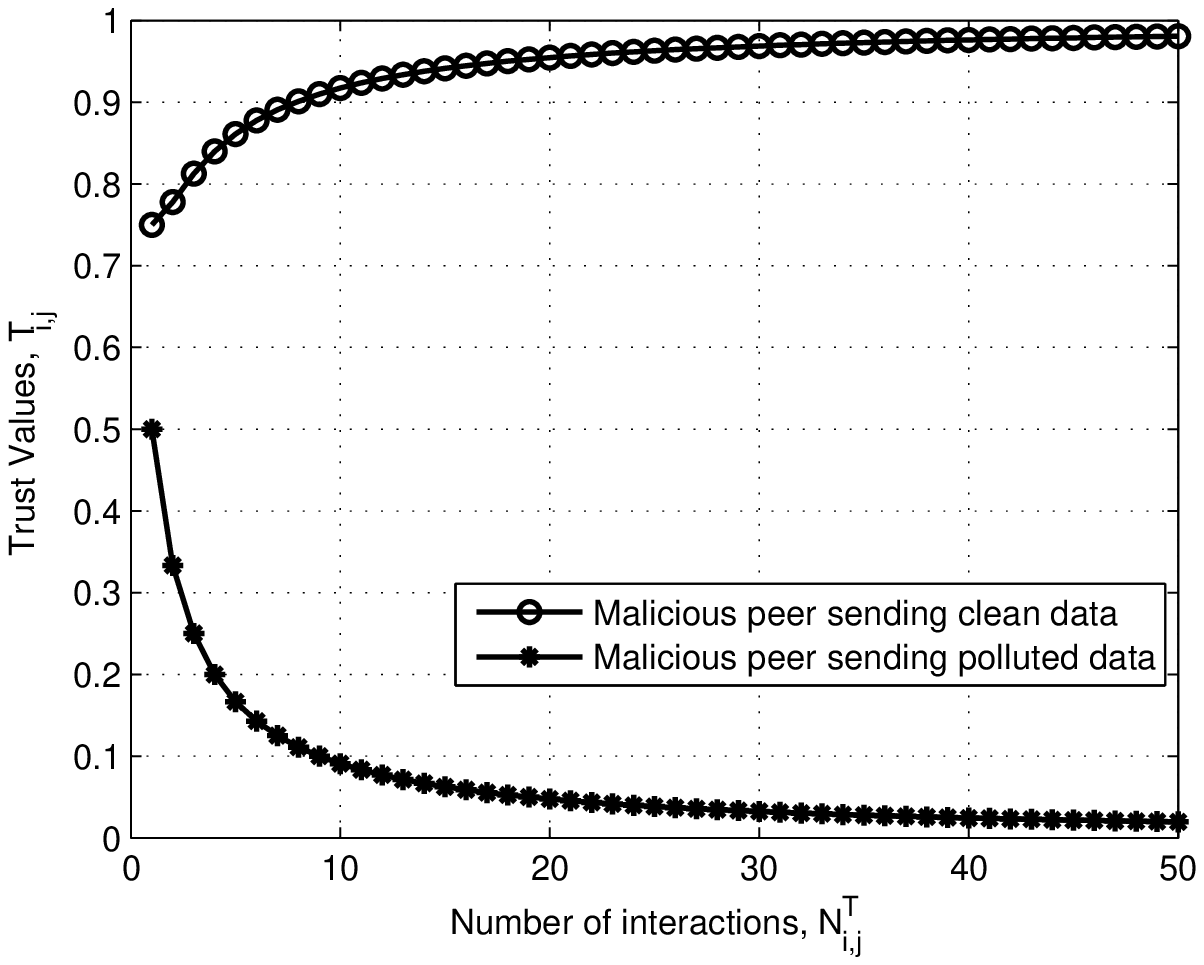}
        \caption{Performance under collaborative attack}
        \label{Fig-Colludeattck}
\end{figure}

\begin{figure}[t]
        \centering
        \includegraphics*[width=10cm]{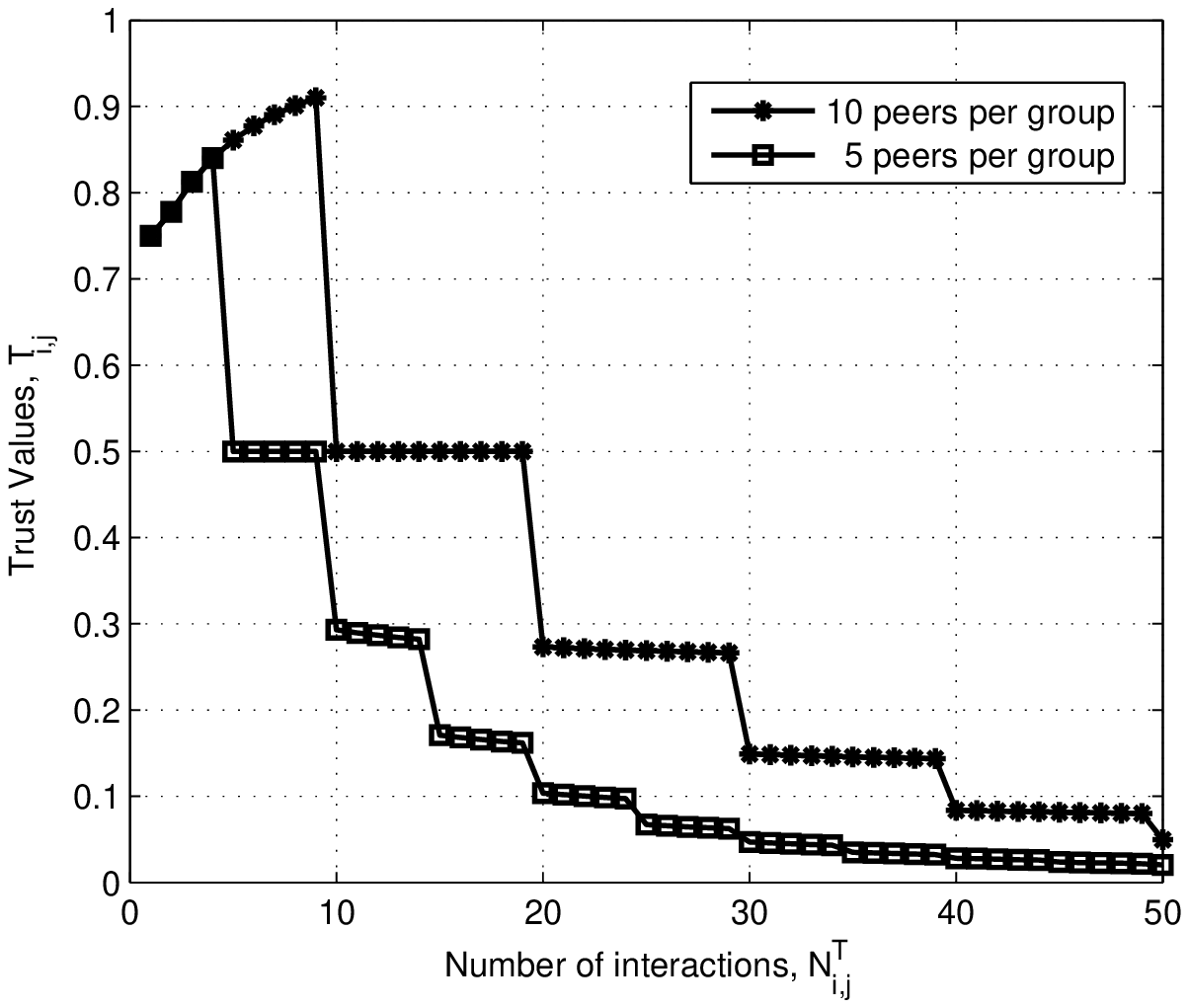}
        \caption{Performance under advanced collaborative attack where malicious peers in the group take turns to attack a target peer}
        \label{Fig-ColludeattckAD}
\end{figure}
In this experiment, we assume there are $10$ malicious peers
carrying out collaborative attack. One of them is chosen to keep
sending polluted data chunks to the target peer, while the remaining
peers send clean data chunks and give high recommendation to the
malicious peer that is chosen to send polluted data chunks. The
trust values of the malicious peer sending polluted data chunks and
a malicious peers sending clean data chunks are plotted in Fig.
\ref{Fig-Colludeattck}. It is observed that the trust values of the
malicious peer sending polluted data drops quickly with the
increasing of the number of the interactions. This is as expected
since that the dynamic confidence factor can effectively reduce the
weight of indirect trust with the increasing of the number of
transactions. When the number of transactions exceeds certain level,
the trust value is dominated by direct trust. The trust value of the
malicious peer that sends clean data chunks is increasing, since
there is no punishment mechanism to prevent peers from giving
misleading recommendations in our system. Actually, it is not
necessary since the target peer can identify the malicious peer that
is sending polluted data chunks, and reject receiving data from it.
At the same time, it can also benefit from receiving clean data
chunks from those malicious peers that pretend to be good peers.

In Fig. \ref{Fig-ColludeattckAD}, we investigate the performance of
the proposed trust management mechanism under the advanced
collaborative attack where malicious peers take turns to attack a
target peer. Two scenarios are considered here. Scenario A
considerers that $10$ peers collude and take turns to attack a
target peer. While scenario B considers a group size of $5$. It is
observed from the figure that the trust values of the malicious peer
in scenario A drop slowly than those of the malicious peer in
scenario B. This is due to the fact that malicious peers in scenario
B attacks the target peer more frequently. In scenario B, each
malicious peer sends a polluted data chunk to the target peer after
sending four clean data chunks, which is similar as the on-off
attack with a on-off ratio of $20\%$. While in scenario A, each
malicious peer sends one polluted data chunk to the target peer
after sending nine clean data chunks, which is similar as the on-off
attack with a on-off ratio of $10\%$. This indicates that the group
size of the collaborative attack matters. Collaborative attacks with
a larger group size are more harmful than those with a smaller group
size.

Besides, comparing scenario A with the $10\%$ on-off attack scenario
given in Fig. \ref{Fig-PDTM}, it is observed that the trust values
of the malicious peer in scenario A drop slowly than those of the
malicious carrying out $10\%$ on-off attack, especially during the
first 30 interactions. This is due to the fact that in scenario A,
the indirect trust value of the attacker is high since other
malicious peers in the group keep giving high recommendations on the
attacker. Thus, the trust value of the attacker drops slow than that
of the $10\%$ on-off attack. However, with the increasing of the
number of interactions, the effect of the indirect trust on the
trust computation decreases due to the role of the dynamic
confidence factor. Thus, the gap between the two cases gradually
diminishes.
\subsection{Experiment 5: The effect of trust values on the number of data requests}
\begin{figure}[t]
        \centering
        \includegraphics*[width=10cm]{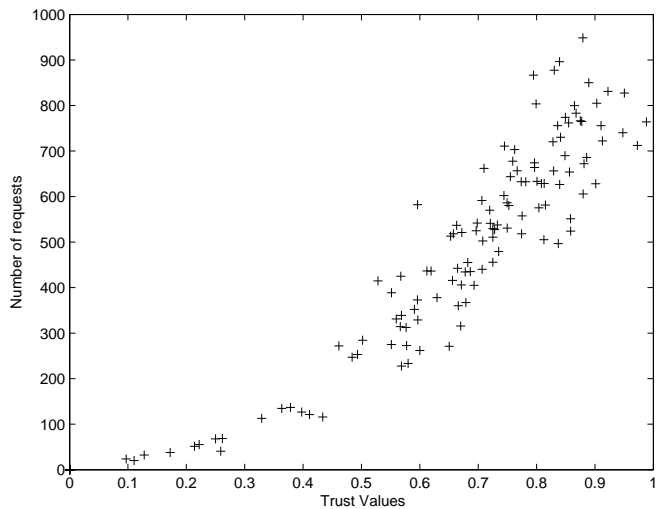}
        \caption{The effect of trust values on the number of data request}
        \label{Fig-ReVsTrust}
\end{figure}
In this experiment, we run our algorithm by implementing an
event-driven script on the real-world testbed PlanetLab
\citep{Planetlab}. We observe the number of data requests at 100
peers with different trust values. The trust values of these peers
is computed by a new peer that just joins the network and does not
have any interactions with others. It is observed from Fig.
\ref{Fig-ReVsTrust} that the peers with large trust values in
general receive more data requests than peers with low trust values.
The peers with trust values lower than $0.5$ receive much less data
requests. This indicates that the proposed trust management
mechanism is quite effective in reducing the peers exposure to
potential malicious peers. On the other hand, it is observed that
the peers with low trust values can still attract some data
requests. These data requests come from the new peers that just join
the network. This is due to the fact that when these peers join the
network, they do not have any interactions with other peers. The
indirect trust plays a dominate role in computing the trust of other
peers. Thus, these new peers are more vulnerable to malicious
attacks. They may send data requests to those peers with low trust
values if they receive misleading recommendations.

\subsection{Experiment 6: Performance comparison between the proposed trust management system and PeerTrust}
\begin{figure}[t]
        \centering
        \includegraphics*[width=10cm]{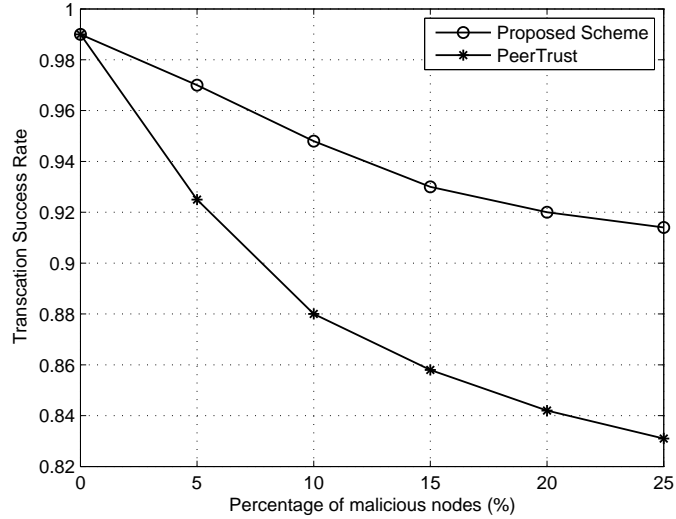}
        \caption{Proposed trust management system vs. PeerTrust}
        \label{Fig-TrustWhole}
\end{figure}

\textcolor[rgb]{0.00,0.00,0.00}{In this experiment, we compare the
performance of the proposed trust management system with that of the
existing PeerTrust given in \citep{LXiong_conf}. For our scheme, we
use the same simulation parameters as in example 3. For each
scenario, we let half of the malicious peers perform persistent
attack, and half of the malicious peers perform on-off attack with a
$20\%$ on-off ratio. The network condition is emulated by randomly
adding a packet loss rate between $0\%$ and $2\%$ at each peer. It
is observed from Fig. \ref{Fig-TrustWhole} that when there is no
malicious peers in the network, the proposed trust management system
performs the same as PeerTrust. However, with the increasing of
percentage of malicious nodes, the advantages of the proposed trust
management system become evident. This is as expected since the
proposed trust management system takes the features of pollution
attack and the network conditions into consideration when it is
designed, while PeerTrust does not.}

\section{Discussion and Future Work}

\subsection{Secure Transmission of Indirect Trust Values}
Due to the distributed nature of P2P multimedia streaming networks,
the unauthorized manipulation of indirect trust values can happen
during the transmission. Thus, it is very important to guarantee
secrecy and integrity of the trust data. This can be achieved by a
PKI-based (Public Key Infrastructure) \citep{AppliedCryptography}
scheme. When a peer $i$ wants to evaluate the trustworthiness of
peer $j$, it initiates an enquiry on the indirect trust value of
peer $j$, and sends its public key together with the enquiry. Then,
the peers who have transaction experience with peer $j$, encrypt
their responses with peer $i$'s public key and sign the responses
with their own private keys. Then, these peers sends the signed
encrypted responses to peer $i$ together with their public keys.
Upon receiving these responses, peer $i$ verifies their signatures
with the attached public keys and decrypts the responses with its
own private key. The fact that the responses are signed with the
responding peers' private keys allows the detection of integrity
violations of the trust values and the authenticity of their
origins. The fact that the trust values are encrypted with peer
$i$'s public key guarantee the confidentiality of the trust data
transmission.

\subsection{Joint Design of Trust Management and Incentive Mechanisms}
Indirect trust plays a significant role in computing the trust value
of a target peer when a peer does not have much interactions with
the target peer. However, without effective incentive mechanisms,
peers have no incentive to cooperate with each other, and thus the
trust records based on other peers' recommendations cannot be
quickly established. Therefore, effective incentive mechanisms
\citep{XKangIncentive,XinJICC} are crucial for the successful
implementation of the proposed trust management mechanism. On the
other hand, trust management mechanism are very important for P2P
streaming systems with incentive mechanisms. Without effective
measures to identify malicious peers, the polluted data chunks could
be disseminated to the whole network more quickly in a P2P network
with incentive mechanisms than that without incentive mechanisms.
This is due to the fact that peers are motivated to upload data
chunks to each other to earn points or monetary rewards in a P2P
system with incentive mechanisms. Without the ability to identify
malicious peers, peers are more likely to forward polluted data
chunks, consequently degrading the performance of the system.
Therefore, trust management and incentive mechanisms should be
jointly designed to defend against both malicious attacks and
selfish users. We leave this as our future work.

\subsection{Tuning of the Parameters}
\textcolor[rgb]{0.00,0.00,0.00}{In the proposed trust model
presented in this paper, there are a lot of tunable parameters. The
value of these parameters plays a significant role in the
effectiveness of the proposed trust model. Thus, how to optimally
choose the values of these parameters is of great importance. In
this paper, we deign these parameters as tunable parameters and let
the users decide these parameters based on their own benefits. This
is due to the fact that different users have different requirement.
It is clear that this design offers great degree of freedom, and
benefits each individual peer. From the perspective of the whole P2P
community, this may not be optimal. Besides, the value of these
parameters may have an impact on the network topology resilience,
the streaming quality, and the network awareness \citep{XinJ}. Thus,
how to choose these parameters such that the performance of the
whole community can be optimized needs to be investigated. We leave
this as our future work.}

\section{Conclusion}\label{conclusions}
In this paper, a trust management system to fight against various
kinds of pollution attacks for P2P multimedia streaming systems are
proposed by exploring the unique features of pollution attacks. A
dynamic confidence factor is proposed to dynamically adjust the
weight of direct and indirect trust in computing the trust, which is
shown to be pretty effective in fighting against the bad-mouthing
attack. Guidelines on how to deign such a dynamic confidence factor
are given, and two specific designs of the dynamic confidence factor
are proposed. Besides, a new direct trust model that is proved to be
resistant to the on-off pollution attack is proposed and
investigated. The performance of the proposed trust management
mechanism under various types of pollution attacks is then
investigated.
Finally, several numerical examples are presented, which show the superiority of the proposed trust management system. 

\section*{ACKNOWLEDGMENT}
\textcolor[rgb]{0.00,0.00,0.00}{We would like to express our sincere
thanks and appreciation to the associate editor and the anonymous
reviewers for their valuable comments and helpful suggestions. This
has resulted in a significantly improved manuscript.}


\end{document}